\documentclass[sigconf]{acmart}

\usepackage{multirow}
\AtBeginDocument{%
  }

\setcopyright{acmlicensed}
\copyrightyear{2025}
\acmYear{2025}

\copyrightyear{2025}
\acmYear{2025}
\setcopyright{cc}
\setcctype{by-nc-nd}
\acmConference[ICAIF '25]{6th ACM International Conference on AI in Finance}{November 15--18, 2025}{Singapore, Singapore}
\acmBooktitle{6th ACM International Conference on AI in Finance (ICAIF '25), November 15--18, 2025, Singapore, Singapore}\acmDOI{10.1145/3768292.3770405}
\acmISBN{979-8-4007-2220-2/2025/11}

\begin{document}

\title{Right Place, Right Time: Market Simulation-based RL for Execution Optimisation}

\author{Ollie Olby}
\affiliation{%
  \institution{Simudyne}
  \country{UK}
  }

\author{Andreea Bacalum}
\affiliation{%
  \institution{Simudyne}
  \country{UK}
  }

\author{Rory Baggott}
\affiliation{%
  \institution{Simudyne}
  \country{UK}
  }

\author{Namid R. Stillman}
\affiliation{%
  \institution{Simudyne}
  \country{UK}
  }
\email{namid@simudyne.com}

\begin{abstract}
Execution algorithms are vital to modern trading, they enable market participants to execute large orders while minimising market impact and transaction costs. As these algorithms grow more sophisticated, optimising them becomes increasingly challenging. In this work, we present a reinforcement learning (RL) framework for discovering optimal execution strategies, evaluated within a reactive agent-based market simulator. This simulator creates reactive order flow and allows us to decompose slippage into its constituent components: market impact and execution risk. We assess the RL agent’s performance using the efficient frontier based on work by Almgren and Chriss, measuring its ability to balance risk and cost. Results show that the RL-derived strategies consistently outperform baselines and operate near the efficient frontier, demonstrating a strong ability to optimise for risk and impact. These findings highlight the potential of reinforcement learning as a powerful tool in the trader’s toolkit.
\end{abstract}



\maketitle

\section{Introduction}
Execution algorithms are a vital tool for trading. They allow market participants to optimally execute large meta-orders by decomposing them into a series of smaller orders and submitting them over a specified time horizon. This is essential for reducing slippage, which is defined as the difference between the expected and actual execution price of a trade. Slippage occurs when the market moves while a meta-order is being executed, incurring a cost to the trader, also known as an execution cost. Slippage, $\zeta$, can be defined as the sum of market risk, $\zeta_{MR}$, and market impact, $\zeta_{MI}$,
\begin{equation}
    \zeta = \zeta_{MR} + \zeta_{MI}
\end{equation}
where market impact is described as the price movement caused by the execution of the order itself, and market risk is characterised as the price movement resulting from exogenous factors. This slippage or execution cost can often be significant, particularly when trying to execute large positions within a short horizon.

Optimising these execution algorithms is therefore key for reducing the adverse cost to a trading firm. Almgren and Chriss presented seminal work on the optimisation of these execution algorithms, where the optimal trade off between risk and reward is defined as a frontier given by the equation \cite{almgren2000optimal}
\begin{equation} \label{eq:frontier}
    \min_x (E(x)+\lambda V(x))
\end{equation}
where $E(x)$ and $V(x)$ are the mean and variance of the transaction cost, respectively. This frontier provides a useful insight into the performance of a particular strategy but does not help find novel strategies without exhaustive and expensive sampling procedures.

The generation of these novel strategies is a challenging task. Typically, these strategies are generated and optimised using backtesting approaches, which, while useful, are inherently limited. Backtests rely on historical data that may not capture the full spectrum of market dynamics, nor react to trades made by the trader agents. This is often a  critical components of market costs. As a result, strategies that appear effective in backtests may perform poorly in live environments.

To overcome the issues synonymous with backtests, we propose using an agent-based model (ABM) for realistic market simulation. ABMs provide the functionality to model the diverse strategies, reactions, and mechanisms of various market participants, generating synthetic data that reflects plausible market conditions. This reactive, synthetic environment provides a controlled, repeatable testbed where different execution strategies can be evaluated for market impact, risk and slippage separately.

\begin{figure}[t]
    \centering
    \includegraphics[width=0.8\linewidth]{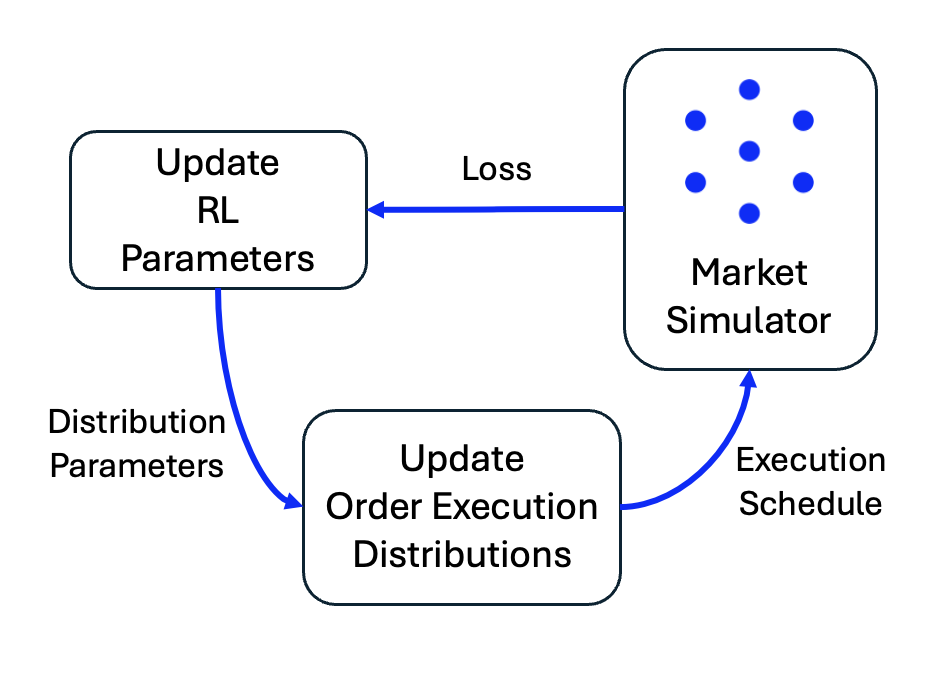}
    \caption{Overview of the proposed framework using the \texttt{Simudyne Pulse} intraday market simulator.}
    \label{fig:high_level}
    \vspace{0.7cm}
\end{figure}

Data-driven models, specifically reinforcement learning (RL) and policy models could be used to generate novel, interpretable and performant execution algorithms. Policy models learn from the simulation environment, iteratively refining their recommendations to efficiently identify optimal strategies. In this work, we use a market simulator to provide a reactive environment which can decompose market impact and risk from slippage for the RL agent to trade in\cite{vytelingum2025agentbasedliquidityriskmodelling}. The RL agent can then learn the impacts of its actions to generate optimal trading strategies. Figure~\ref{fig:high_level} shows a high level representation of the framework outlined in this paper.
By using an RL agent to find optimal parameters for a known trading algorithm template, we ensure that the output is sufficiently interpretable. This interpretability is key because the EU AI Act of 2024 dictates that, if using a high-risk AI system (such as is typically used in trading decisions), any person subject to the output of the AI has the right to clear and meaningful explanations of the role of the AI system \cite{EU_AI_Act_2024}.

In this study, we first compare RL algorithms that consider different distribution of orders, such as a uni-modal or bi-modal distribution. We also consider how the RL algorithms perform using different metrics of market costs, such as slippage, risk, and impact, which we can directly calculate using our market simulator. Secondly, we show how our results integrate with classical frameworks for quantifying execution costs. Specifically, we compare the RL devised execution algorithms with the Almgren and Chriss efficient frontier \cite{almgren2000optimal}

\section{Related Works}\label{related_work}

\subsection{Agent-Based Simulators}
One popular model for recreating the price impact without the limit orderbook is the Chiarella model \cite{Chiarella01102002}. It models the supply and demand of different types of traders and aggregates them into a demand signal which relates linearly to asset price via Kyle's lambda. However, this model does not include order matching and limit orderbook dynamics, which are required for accurate market impact modelling.

Recent years have seen a new class of agent-based models (ABMs) that explicitly model the market auctions with limit orderbooks and reproduce realistic market behaviours. Some of the most popular models build on the zero intelligence model \cite{Farmer2005, Sunder1993, Preis2006} and ABIDES \cite{byrd2019abideshighfidelitymarketsimulation, Coletta_2022}.

Many of these models use rule-based agents. However, these require a vast set of calibrated parameters. By using zero (or minimally) intelligent models and placing orders sampled from a distribution, the order flow can be recreated \cite{Farmer2005}. For example, the stochasticity of the order flow can be fitted to historical data to create a simple simulator that does not rely on explicit rules of order execution. The zero intelligence model can explain properties of a financial market, such as the variance of the bid-ask spread and of the price diffusion rate despite its simplicity. However it also lacks the complexity of a more realistic market simulator.

ABIDES is a popular multi-agent framework which incorporates three types of trader behaviour: mean reversion, trend following and noise. ABIDES enables the simulation of tens of thousands of trading agents interacting with an exchange agent. This framework uses a greedy parameter to model market impact however this is not directly calibrated \cite{byrd2019abideshighfidelitymarketsimulation}.

\subsection{Optimal Trade Execution}
Almgren and Chriss introduced one of the most influential mathematical models for optimal trade execution \cite{almgren2000optimal}. They presented a framework to calculate the transaction cost of executing a large meta-order. They compared different strategies with different time-dependent rates and found a frontier of optimal strategies that minimised execution cost for each value of variance. A strategy that tries to execute the total volume immediately will have maximum cost and a low variance, while a risk-neutral strategy will achieve the lowest average cost at the expense of increased variance of execution. This frontier of optimal execution algorithms (Equation \ref{eq:frontier}) is defined as a smooth curve outlining the set of trading strategies for which no alternative strategy offers both a lower expected cost and lower risk.

Equation \ref{eq:frontier} relates the frontier to $\lambda$, which can be viewed as a measure of risk aversion or how much we penalise risk relative to cost. When $\lambda = 0$ this represents a so-called ``na\"ive strategy" where the execution happens uniformly over the full trading day. This is the most efficient strategy on the frontier, where moving in either direction to reduce or increase the risk will result in a more expensive trade. Conversely, when $\lambda \to -\infty$, this represents a strategy where the trader executes all available volume as quickly as possible. This has very low market risk but can have a high execution cost due to adverse market impact created by a quick execution of a large order. When $\lambda \to \infty$, this represents a strategy where the trader waits to execute all available volume until the end of the day. This has a very high market risk but can have a low market impact, as execution is deferred until favorable market conditions arise.

Value at risk (VaR) is a risk measure that estimates the maximum expected loss a portfolio may experience over a given time horizon, with a specified level of confidence. For a trading strategy $x$, Almgren and Chriss defined VaR as the level of transaction cost that will not be exceeded with probability $p$. They assumed that price dynamics followed an arithmetic Brownian motion. This means trading costs are normally distributed with a known mean and variance. The confidence level is then determined by $\lambda_p$, representing the number of standard deviations from the mean. Consequently, VaR can be expressed as

\begin{equation}
    \mathrm{VaR}_p(x) = \lambda_p \sqrt{V(x)} + E(x).
\end{equation}

Thus, with probability $p$, the trading strategy $x$ is not expected to incur trading costs exceeding $\mathrm{VaR}_p(x)$. A strategy $x$ is considered efficient if it achieves the minimum possible VaR for the given confidence level $p$, shown on the efficient frontier as a tangent at the specified confidence level, $\lambda_p$.

Building on analytical models like Almgren-Chriss, later work explored data-driven approaches for optimal trade execution. For example, Nevmyvaka et al. presented one of the first applications of RL to the problem of optimal trade execution \cite{nevmyvaka2006reinforcement}. The authors developed a custom RL algorithm that learns policies for optimally placing limit orders over a fixed execution horizon. Their work established RL as a powerful tool for optimising trading algorithms.

De Meer Pardo et al. proposed a modular framework for using RL to build optimal trading algorithms \cite{yao2024reinforcementlearningagentbasedmarket}. They emphasised the importance of a realistic simulation environment. This framework decomposes the environment into interchangeable modules. The authors demonstrate the framework by augmenting a TWAP benchmark with an RL agent to adjust order volumes within a fixed schedule. Their experiments show modest improvements over the TWAP baseline.

Hafsi and Vittori presented a reinforcement learning-based framework for optimal trading algorithm discovery, using the ABIDES market simulator as the training environment \cite{hafsi2024optimalexecutionreinforcementlearning}. They employed deep Q-networks, with a reward function that applied penalties for both order book depth consumption and unexecuted order volume. Their results showed that the RL agent outperforms traditional strategies such as TWAP, particularly in terms of execution consistency and reduced market impact.

Lin and Beling applied proximal policy optimisation to train an agent for optimal trade execution, where the action space corresponds to the number of shares to execute at each decision point \cite{10.5555/3491440.3492067}. The authors introduced a reward function that provides feedback only at the end of each episode, based on the agent’s performance relative to a TWAP benchmark. The policy is modeled using either a multilayer perceptron or a long short-term memory network, enabling the agent to learn directly from raw level-2 limit order book data without the need for manual feature engineering.

\section{Methodology} \label{sec:ABM}
In this section, we outline the details of our agent-based market simulator and our RL trading agent. We explain how we employ RL to optimise an execution algorithm using a realistic market simulator. Finally, we generate the Almgren-Chriss efficient frontier using our market simulator to explore the efficiency of the RL devised strategy.

\subsection{Market Simulator}
All tests are run using the \texttt{Simudyne Market Simulator}. It provides a realistic market environment that generates statistically accurate market behaviour.

Our market simulator has three main components: the \textit{exchange} module which matches the protocols and behaviour of a specified exchange, the \textit{agents} who have certain behavioural characteristics, and the \textit{calibration} module which tunes the parameters of the overall simulator to replicate behaviour of a specific asset on a specific historical date. The simulator recreates the statistical properties of the market using a network of trader agents who are connected to the exchange. At each step, trader agents send messages to submit limit or market orders and amend or delete queued limit orders. The Exchange then sends execution reports back to the Trader as well as level-2 market data.

\subsubsection{The Exchange}
The exchange determines the protocol by which the limit orders are queued, matched and priced. The exchange's matching engine will attempt to match each order sequentially. When a sell limit order is less or equal to a buy limit order, a trade occurs and is priced at the earlier order and completely matched orders are removed from the LOB. Unmatched or partially matched limit orders are queued on the LOB. A market order is immediately matched at any price and unmatched market orders are canceled.

The exchange also dictates how the market reacts to trades. There are different approaches to modelling price impact \cite{Bouchaud_Bonart_Donier_Gould_2018}. Here, we adopt an empirically derived approach, which we refer to as the aggregate impact function. We calculate the price impact over a time scale instead of trade-by-trade. Each second we aggregate the orderflow imbalance (the mid price change due to a period of excess demand $Q$). We then relate mid-price change to the orderflow imbalance that caused them. A positive impact is generated by excess demand and a negative impact is generated by excess supply. This is because market risk should cancel out on average. We then fit $f_{mi}$ to the observed excess and market impact data using a least squares regression. The aggregate impact function is defined as 
\begin{equation}
    f_{mi}(Q) = \lambda Q^\gamma
\end{equation}
where $Q$ is the excess demand, and $\lambda$ and $\gamma$ are both calibrated parameters. 

Transient market impact is the sum of market impact across many trades, it is therefore not possible to isolate the impact from a single trade. Therefore, rather than attempt to estimate the impact of an individual trade we use the aggregate impact function where we assess the impact over a time period instead. Additionally, we take a weighted approach to fitting our aggregate-impact function, where the weights are set to the square root of the number of observations for a robust estimation that does not penalise information from less frequent, larger orderflow imbalances.

\subsubsection{The Agents}
The rate of order submission (both market and limit) can be estimated by assuming a Poisson model \cite{Farmer2005}. Each minute, we estimate a constant arrival rate of limit and market orders and scale to the arrival rate at each step, $t$, as $\alpha(t)$ and $\mu (t)$, respectively.

We then sample limit order prices with a probabilistic model of depth, volume, duration and the volume of a market order, conditional on the state of the LOB. At each time step, $t$ (1 millisecond resolution), an agent has a probability $\alpha(t)$ of submitting a limit order or $\mu(t)$ a market order based on the joint probability distributions. The traders are characterised by alpha strategies that decide whether to buy or sell based on the market state. We use work devised by Majewski et al. \cite{majewski2018coexistencetrendvaluefinancial} and our own work to devise a new set of Chiarella traders with different behaviour types. These include fundamental, momentum (high frequency and low frequency), noise traders, market makers and deep liquidity providers. Given the demand $D(t)$ from the Chiarella trader, the probability that this trader submits a limit order is $\alpha(t) D(t)$ and a market order is $\mu(t) D(t)$. We detail the calibration process of the probability distributions and the Chiarella model in the next section.

\subsubsection{Calibration}
In this work we use data from the Hong Kong stock exchange and clearing house as our test market. Specifically, we consider the Hang-Seng Index Futures ticker for the dates of 11-12th September 2023. We rebuild the limit orderbook using the market data as well as a list of all limit orders submitted. The market orders are then inferred from the list of executed limit orders.

Parameters $\alpha$ and $\mu$ (rates of order submission for limit and market orders) are calibrated from the empirical data. We estimate the conditional joint probability function of duration, depth and size given a market state, which is defined as the time the order is submitted and the bid/ask spread just before order submission. We draw a random sample from a subset of all historical orders, conditioned on the bid-ask spread and time from market open, to get a depth, volume and duration whenever a trader submits a limit order. The submission of market orders is calibrated to historical market order volumes, given the submission time and bid-ask spread at submission.

Finally, we need to calibrate the Chiarella model parameters. These parameters dictate the specific behaviours and effective numbers of each type of trader (such as fundamental, momentum and noise). This is a simple optimisation problem where we minimise the distance between the historical market and the simulated market. This distance is quantified by the sum of the distances between a set of measurable stylised facts for the historical and the simulated data. We consider the following stylised facts:
\begin{itemize}
    \item Arrival rate of limit and market orders
    \item Distribution of spread
    \item Distribution of 1s, 60s returns and absolute returns
    \item Autocorrelation of 1s and 60s returns and absolute returns
    \item The persistence of orderflow (the auto-correlation of buy/sell signals)
\end{itemize}

The calibration methods and specific parameters of the \texttt{Simudyne Market Simulator} is discussed in more detail in \cite{vytelingum2025agentbasedliquidityriskmodelling, stillman2023deepcalibrationmarketsimulations}. Here, we use a surrogate model calibration approach, where the final calibration parameters are shown in Table~\ref{tab:pivot_model_params}.

\begin{table}[htbp]
\centering
\caption{Calibrated parameter values for \texttt{Simudyne Pulse}. The parameters include the Chiarella model parameters for the low frequency momentum traders ($\alpha_L$, $\beta_L$, $\gamma_L$) and the high frequency traders ($\alpha_H$, $\beta_H$, $\gamma_H$), market coupling terms ($\kappa$, $\sigma$), and market impact terms (Market $\gamma$, Market $\lambda$). All values are reported to three significant figures. Further explanation of the parameter values can be found in \cite{vytelingum2025agentbasedliquidityriskmodelling}.}
\label{tab:pivot_model_params}
\begin{tabular}{ccc}
\toprule
\textbf{Parameter} & \textbf{2023-09-11} & \textbf{2023-09-12} \\
\midrule
$\kappa$ & 0.0456 & 0.0495 \\
$\alpha_L$ & 0.000173 & 0.000173 \\
$\beta_L$ & 7.77 & 5.81 \\
$\gamma_L$ & 4.84 & 4.40 \\
$\alpha_H$ & 0.981 & 0.981 \\
$\beta_H$ & 1.147 & 1.967 \\
$\gamma_H$ & 168000 & 157000 \\
$\sigma$ & 0.219 & 0.0368 \\
Market $\gamma$ & 0.707 & 0.705 \\
Market $\lambda$ & 0.266 & 0.257\\
\bottomrule
\end{tabular}
\end{table}

\subsubsection{Transaction Cost Calculation}
Transaction cost or slippage refers to the difference between the expected price of a trade and the actual price paid during the execution of a meta-order. This value can be separated into market risk and market impact. Market impact is the movement of the market specifically due to the execution of the meta order. Market risk refers to the price fluctuations due to exogenous factors, whether or not the meta-order is executed.

One novelty of our market simulator is its model of market impact and the resulting reactivity of the market with respect to trades. This allows us to decompose execution cost (slippage) into market risk and market impact. Slippage $\zeta$ is calculated as

\begin{equation}\label{eq:slippage}
    \zeta = \frac{\sum_t p_t^E v_t^E}{v_T^{E}} - p_R
\end{equation}
where $p_t^E$ are the executed trade prices, $v_t^E$ are the executed volumes at each step, $v_T^{E}$ is the total amount of volume executed each day and $p_R$ is the reference price.

\begin{figure*}[t]
    \centering
    \includegraphics[width=0.9\linewidth]{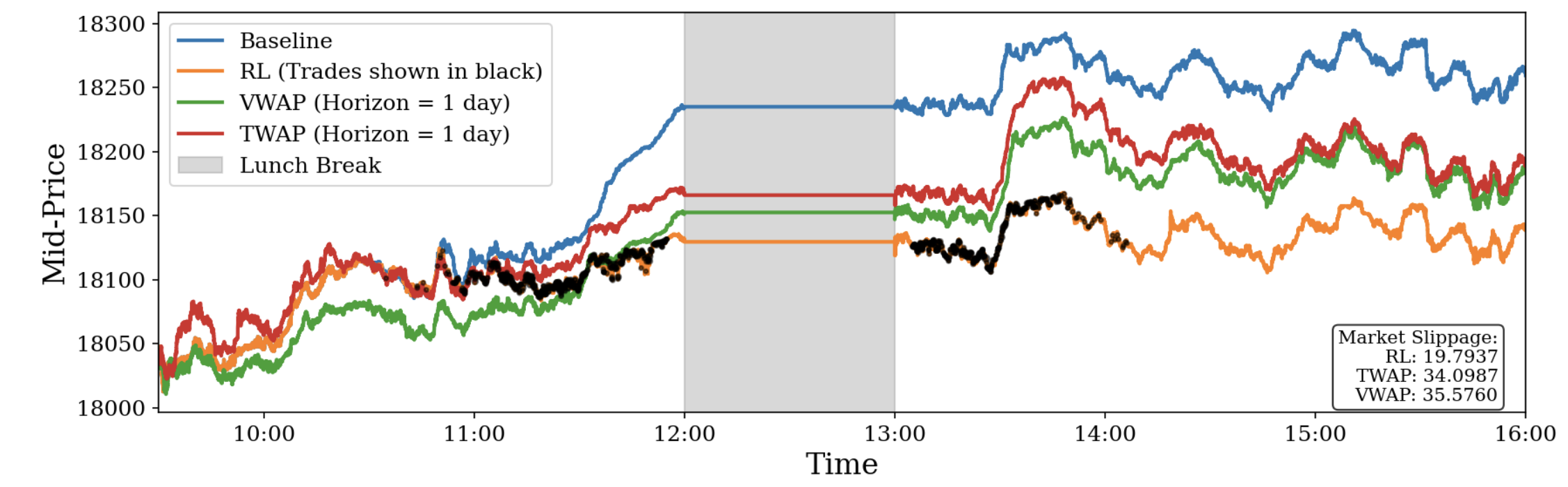}
    \caption{Price paths showing one MC run for the RL learned bi-modal un-bounded execution distribution and the two baselines (TWAP and VWAP). For the RL devised strategy the position and size of the trades are shown in black. The values for slippage are averaged over ten Monte Carlo runs. This is based on data from the Hang Seng futures index from 12th of September 2023}
    \label{fig:price_paths}
\end{figure*}

Using our market simulator to run both a baseline and a counterfactual, we can directly measure the market impact, $\zeta_{MI}$, and market risk, $\zeta_{MR}$. This is done with each trader using the same seed and using the same exogenous signal $V_t$ (which is used by our fundamental traders, as in \cite{majewski2018coexistencetrendvaluefinancial}),

\begin{align}
    \zeta_{MR} &= \frac{\sum_t p_t^B v_t^E}{v_T^{E}} \\
    \zeta_{MI} &= \frac{\sum_t (p_t^E - p_t^B) v_t^E}{v_T^{E}}
\end{align}
where $p_t^B$ are the simulated prices in the baseline. Our simulator allows for a baseline strategy to be generated where actions from the trader agent in question are not considered. On a new price path, but with the same seed, trades are then made by our trader agent to generate the counterfactual price path. From these price paths, the risk and impact of certain trades and positions can be ascertained. Figure \ref{fig:price_paths} shows how the baseline and counterfactual price paths differ and the impact of trades made by the trader agent.

\subsection{Reinforcement Learning}

The RL framework trains a continuous policy gradient model to find the mean, $\mu_k$, and standard deviation, $\sigma_k^2$, for a uni-modal or multi-modal distribution of the time of orders, given the $k^{\text{th}}$ mode of the distribution. The success of each experiment is determined by using our reactive market simulation testing environment and a specified loss function. 

\begin{figure}
    \centering
    \includegraphics[width=0.9\linewidth]{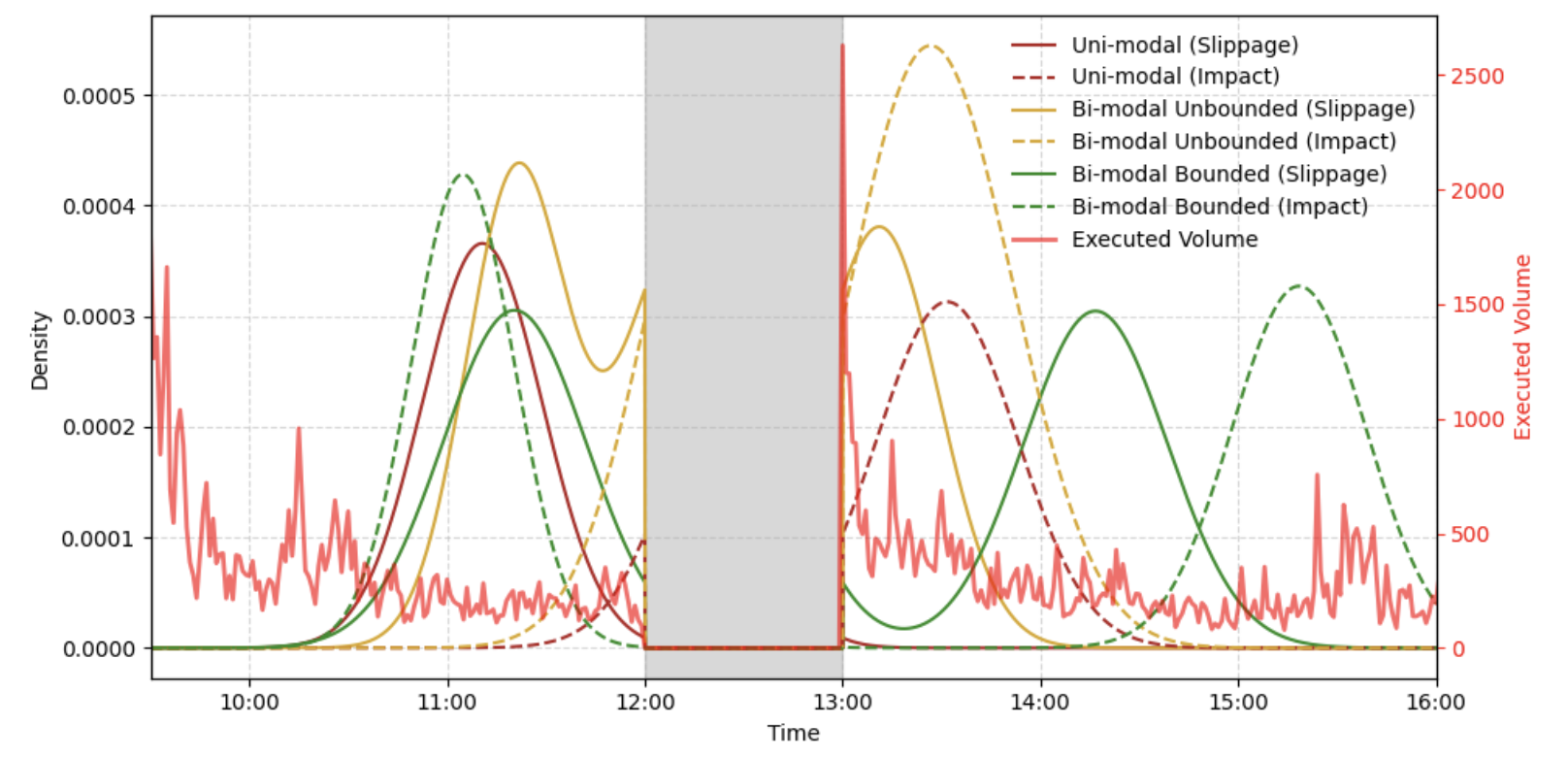}
    \caption{Plot showing the three different distributions found by the RL trading agent for each objective function. The plot also shows the aggregate volume over 1 minute periods in red. The lunch period is shown as a shaded region.}
    \label{fig:distributions}
\end{figure}

The specific schedule of orders is sampled from the distribution with $K$ modes. Figure \ref{fig:distributions} shows the discovered distributions for the three modal distributions that we considered. We sample $N/K$ orders from each distribution, where $N$ is the number of total executed orders (1000), each with a size of $n=\frac{5000}{N}$ units for a total order volume of 5000 units. We define our sets of sampled order schedules as
\begin{align}
\omega_m =n\; & \left\{ t_i^{(k)} \sim \mathcal{N}(\mu_k, \sigma_k^2)\;\middle|\;
 \text{for } k \in\{1,\dots,K\}, \space i=1,\dots,N \space \right\}
\end{align}
where $t$ is the time of execution, and $\mathcal{N}(\mu_k, \sigma_k^2)$ is the normal distribution with mean $\mu_k$ and variance $\sigma_k$. These values are defined as the action of the policy model, $a_k=(\mu_k, \sigma_k)$, for the $k^\text{th}$. A total of $N$ orders are sampled per distribution, each with a fixed size, $n$, and executed at the corresponding timestamps $t_i^{(k)}$.

The RL algorithm used here is very similar to a multi-armed bandit model \cite{bouneffouf2020survey}. The name, multi-armed bandit comes from building models to try and beat slot machines (colloquially known as one-armed bandits) by learning which arms generate certain reward patterns. Similarly, in our setting, each decision to place an order can be viewed as pulling an arm of the slot machine, where the model aims to learn the optimal strategy (policy) for slicing a parent order into the market. Future work will seek to extend this to include market context, such that our RL algorithm becomes a contextual bandit model \cite{cannelli2023hedging, bouneffouf2020survey}. 

The RL model interacts with our simulated trading environment to learn the policy $\pi_{\theta}$. The loss is determined by executing the order schedule through our market simulator and computing the market slippage, impact and risk. The policy model is a lightweight feed-forward neural network, where the output are the parameters for a uni-modal or multi-modal Gaussian distribution. As discussed, we do not pass context to the RL model, which we leave as future work. Hence, our work demonstrates the benefit of combining a reactive market simulator with even a relatively simple RL model for order execution. The network structure is defined as:

\begin{align}
h &= \mathrm{ReLU}(W_s x + b_s) \\
\mu_k &= \frac{\tanh(W_k^\mu h + b_k^\mu)+1}{2} \\
\sigma_k &= \frac{\tanh(W_k^\sigma h + b_k^\sigma)+1}{2}
\end{align}
where $x$ is the input state vector (which we set as a constant dummy state to account for the lack of context in the model), where $W$, $b$ terms are learned parameters, and where $k$ represents the modes of the uni-modal or multi-modal distribution. The $\tanh$ function was used to clip the values between 0 and 1 (which were then rescaled to represent the whole trading day).

During training, for every episode, the model samples several sets of schedules of orders ($\omega_m$) to create a batch of schedules, where $m$ defines the schedule index within a batch. Each schedule in the batch is evaluated in the market simulator and the reward is the average market slippage across each batch of schedules, where all samples are from the same action-created distribution. We denote the policy, $\pi_\theta(a_m)$, as representing the probability of selecting actions $a_m$, for schedule $m$, where actions are individual order schedules built from execution times, $\omega_m$ and sampled from the multi-modal normal distribution given by the RL model. 

By taking the average log probabilities, the policy loss for each $m^\mathrm{th}$ schedule can be computed as
\begin{equation}
    \mathcal{L}_m = \log\pi_\theta(a_m) \times \hat{r}_m+\beta \cdot \mathcal{H}[\pi_\theta(\cdot|s_m)] 
\end{equation}
where $ \mathcal{H}[\pi_\theta(\cdot|s_m)]$ is the entropy of the state $s_m$, and $\beta$ is the entropy coefficient. We also included entropy in the loss function to help encourage exploration of the space and reduce the effects of the $\log \pi_\theta(a_m)$ on the $\sigma_k$ (standard deviation) parameter of the output. Without the entropy term the standard deviation is pushed towards zero, causing the policy to become overly deterministic too early. Entropy regularisation counteracts this by encouraging higher uncertainty, preventing the agent from converging to a sub-optimal strategy. Finally, $\hat{r}_m$ is the reward function, which is defined as:
\begin{equation}
    \hat{r}_m = -\overline{\mathrm{loss}}.
\end{equation}
The loss is averaged across all samples within the batch to get an average loss for that specific distribution.

The policy, $\pi_\theta$, is then updated using the REINFORCE algorithm \cite{williams1992simple}. We compute gradients of the expected return with respect to the policy parameters and adjust them to decrease the overall loss function, in our case market slippage. REINFORCE was used over other more modern RL techniques due to its ease of implementation and interpretability. The model in this work serves as a baseline for future work to explore the gains of using other RL methods such as PPO, DQN and GRPO \cite{ghasemi2025comprehensivesurveyreinforcementlearning}.

The RL was trained with a learning rate of $\mathrm{lr}=0.01$, early stopping after 15 episodes with no decrease in loss, the entropy coefficient was set to $\beta=2.0$ but would decrease after 10 episodes to $\beta=0$ and we used a batch size of 10 with 50 Monte Carlo samples each. To get these values, we used a coarse grid-search across learning rates and architecture sizes. However, we found that for all configurations, the model reached convergence within the first twenty episodes. Therefore, we used values which would converge faster to increase the total number of experiments we could run, and to focus our compute time on exploring more points within the efficient frontier.

\section{Results and Discussion}\label{sec:results}

In this section we discuss the results of using an RL agent to find an efficient execution algorithm. First, we compare the RL agent's ability to generate three different trade-time distributions. These are compared against two baseline strategies, TWAP and VWAP. We then compare the devised execution algorithm to the efficient frontier proposed by Almgren and Chriss. The RL agent training is run on data from the Hong Kong stock exchange, specifically, the Hang Seng index (expiring in September) from the 11th of September 2023, the chosen schedule is then tested on data from the same index and exchange but from the 12th of September 2023.

\begin{figure*}[t]
    \centering
    \includegraphics[width=0.85\linewidth]{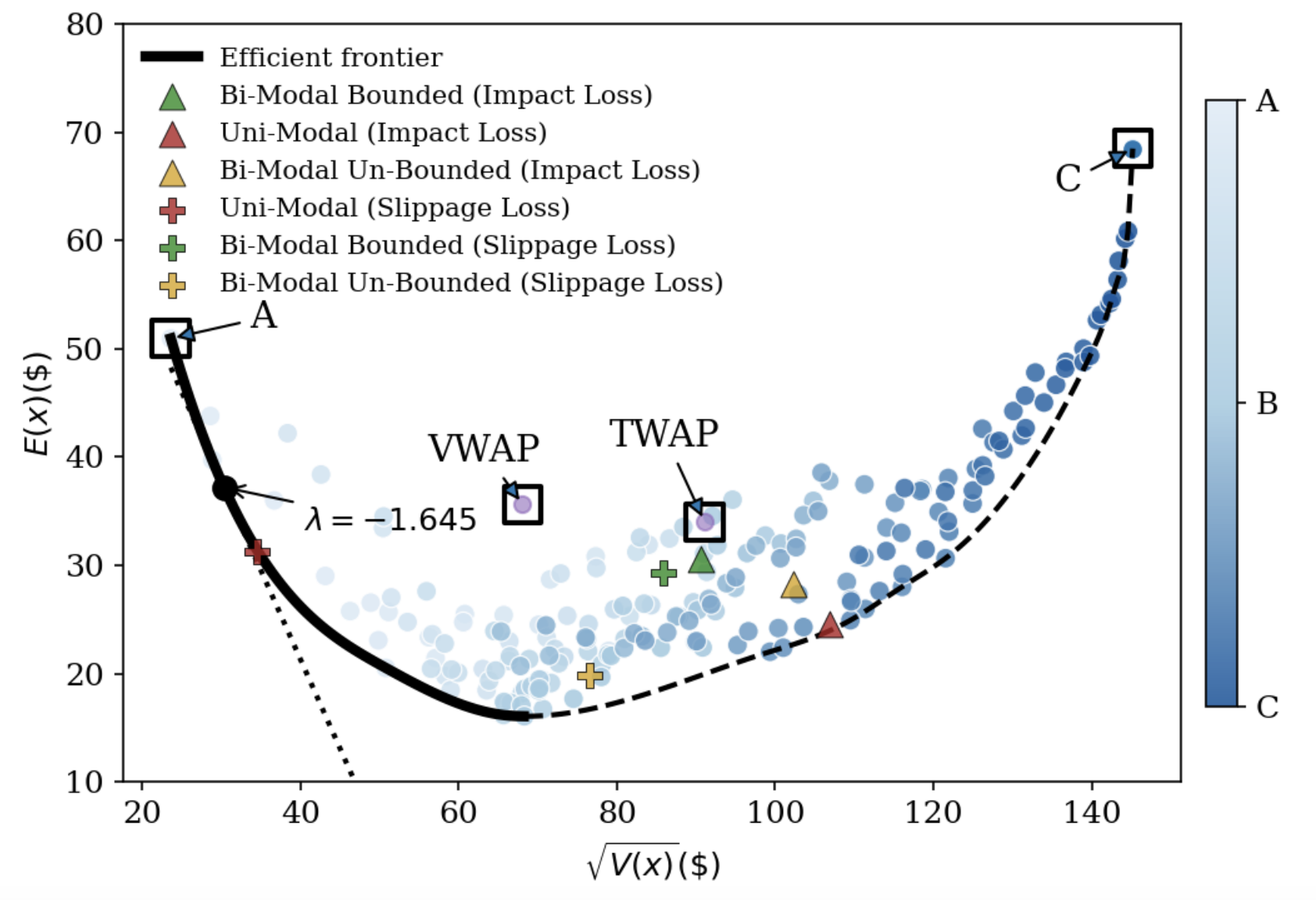}
    \caption{Plot showing the efficient frontier, defined by a comprehensive sample of TWAP strategies with time-differing start and end points. The three reference strategies are shown, where A is a risk-adverse strategy where all the available volume is executed as fast as possible, B is a risk-neutral strategy where the volume is executed evenly over the full day and C is a risk-seeking strategy where all the volume is executed at the end of the day. The dashed part of the frontier shows the sub-optimal part of the frontier, where similar transaction cost are incurred but with a higher amount of market risk. This region is defined as being to the right of the inflection point or where $\lambda =0$. The RL devised strategies are shown alongside the baselines. The tangent line at $\lambda = -1.645$ is shown, this shows the plane of strategies for an optimal VaR at a 95\% confidence level.}
    \label{fig:Frontier}
\end{figure*}

\subsection{Distribution Modes}

We tasked the RL agent to find the optimal parameters that describe three types of order-time distributions: uni-modal unbounded, bi-modal unbounded and uni-modal. Bounded refers to having each mode of the bi-modal distribution constrained to the morning session and afternoon session respectively. We used two different loss objectives; market impact ($\mathrm{loss} = \zeta_{MI}$) and market slippage ($\mathrm{loss} = \zeta$).

The results are shown in Table \ref{tab:distributions}, with the discovered distributions shown in Figure \ref{fig:distributions} and an example of resulting price paths for a bi-modal unbounded distribution shown in Figure \ref{fig:price_paths}.

Under the slippage loss function, results indicate similar market impact values across RL and baseline strategies. This is likely due to the relatively small meta-order size, which inherently limits market impact and thus reduces the potential gains from optimizing it. However, when we look at the transaction costs (slippage) of the proposed strategies, we see improvement for all distributions over the baselines and a considerable improvement for the unbounded distribution. This benefit comes from the reduction of risk, as it is likely the RL is using risk as a shortcut to find more optimal strategies by simply reducing the risk which is comparably simpler to do than reduce the impact.

When optimizing directly for market impact, a novel feature of our simulator, the RL agent consistently reduced impact values compared to both baselines and the slippage-optimised strategies. However, this came at the cost of increased market risk, particularly evident in the bi-modal and unbounded strategies. This trade-off is expected: optimizing impact does not penalise risk, whereas slippage explicitly balances both. Despite this, the RL strategies trained to minimise impact still often resulted in better slippage than the traditional baselines, suggesting secondary benefits of impact-focused learning.

\begin{table}[t]
    \centering
    \renewcommand{\tabcolsep}{4pt}
    \footnotesize
    \caption{Comparison of risk, impact, and slippage across execution strategies. Based on a sell meta-order size of 5000, executed over 1000 orders. Distribution parameters are optimised by an RL agent over 50 episodes; values in the table are averaged over 10 Monte Carlo runs from the test data. Distance refers to the Euclidean distance from the efficient frontier.}
    \label{tab:distributions}
    \begin{tabular}{p{0.9cm}p{2.3cm}ccccc}
        \toprule
        \textbf{Loss} & \textbf{Strategy} & \textbf{Risk} & \textbf{Impact} & \textbf{Slippage} & \textbf{Distance} \\
        \midrule
        \multirow{3}{*}{Slippage} 
            & Uni-modal                  & -11.24 & 42.50 & 31.26 & 0.13 \\
            & Bi-modal (Bounded)         & -14.42 & 45.40 & 30.98 & 10.34 \\ 
            & Bi-modal (Unbounded)       & -28.75 & 48.54 & 19.79 & 3.07 \\
        \cmidrule(lr){1-6}
        \multirow{3}{*}{Impact} 
            & Uni-modal                  & -6.84 & 31.44 & 24.60 & 0.62 \\
            & Bi-modal (Bounded)         & 0.85 & 29.69 & 30.54 & 10.36 \\
            & Bi-modal (Unbounded)       & -2.67 & 30.89 & 28.22 & 5.38 \\
        \hline\hline
        \multirow{2}{*}{Baseline} 
            & VWAP                       & -5.33 & 40.90 & 35.58 & 19.54 \\
            & TWAP                       & -8.47 & 42.57 & 34.09 & 13.60 \\
        \bottomrule
    \end{tabular}
\end{table}

Figure \ref{fig:distributions} shows the learned distributions alongside the intraday volume (aggregated over one minute) for the training data. From this, we can interpret what the RL is doing in its optimisation. A common approach to minimise market impact is to essentially hide large orders in periods of heightened volume, when looking at the distributions, we can see the peaks for the impact objective function are centered around the post lunch volume spike. The bi-modal unbounded strategy places both modes near this high-volume region, explaining its performance advantage. In contrast, the bounded variant consistently exhibits less structured distributions, likely due to stricter constraints making optimization more difficult.

When optimizing for slippage, the agent learns to prioritise early execution to reduce risk exposure. This is reflected in the forward shift of distribution peaks earlier in the trading day. For bi-modal unbounded strategies, one mode remains in the high-volume region to mitigate impact, while the other shifts earlier to address risk, demonstrating the agent’s ability to balance multiple objectives simultaneously.

\subsection{RL on the Efficient Frontier}
Next, we compute the Almgren-Chriss efficient frontier for the test day using 10 Monte Carlo samples of simulations across 190 different time weighted execution strategies. This is shown in Figure \ref{fig:Frontier}. The three reference execution strategies can be characterised as follows. Strategy A seeks to execute all available units immediately, minimising risk but sacrificing a less optimal transaction cost. Strategy B adopts a time-weighted approach, distributing executions evenly throughout the trading day to balance market impact and risk. Strategy C defers all executions until the end of the day, concentrating activity in the closing period to potentially benefit from end-of-day pricing dynamics despite increased market risk. We also plot the RL generated execution strategies for both objective functions (market impact and slippage).

Strategies that are closest to the frontier are deemed more efficient as they efficiently balance risk and impact. We see through Figure \ref{fig:Frontier} and Table \ref{tab:distributions}, that the RL is generating strategies which are closer to the frontier than the baselines. 

When examining the market impact objective function, we see that in order to get an advantageous market impact, the RL sacrifices market risk. This is shown by the RL strategies being close to the sub-optimal portion of the frontier (shown as dashed lines). This is due to the impact function not rewarding a reduction of risk.

We expect the slippage objective to produce the best results on the frontier and this is what we see. We expect this because the frontier requires balancing both risk and impact, and this is equivalent to balancing for slippage (which weights both risk and impact equally). The bi-modal unbounded strategy for slippage is close to point of optimal efficiency ($\lambda =0$). This point is where any increase in risk will not yield better cost, effectively the most optimal expense of risk for minimal cost.

Finally, the uni-modal strategy optimised for slippage lies close to the tangent at $\lambda = -1.645$, corresponding to the optimal Value-at-Risk (VaR) at a 95\% confidence level, as defined by Almgren and Chriss \cite{almgren2000optimal}. This strategy therefore has a 95\% confidence level that the strategy will not exceed an optimal level of VaR (shown by the tangent). This reinforces the conclusion that even with relatively simple loss functions, RL can produce near-optimal execution strategies when appropriately guided by objectives and simulators that reflect market realities.

\section{Conclusion}\label{sec:conclusion}

In this work, we have shown that reinforcement learning (RL) can be effectively applied to the problem of optimal trade execution. Our RL agent, trained in a realistic agent-based simulation, produced execution strategies that outperform traditional baselines such as TWAP and VWAP in terms of slippage and proximity to the efficient frontier.

When benchmarked against the Almgren-Chriss efficient frontier, the RL-derived strategies consistently lie close to the frontier. This suggests the agent is capable of learning strategies that efficiently balance risk and impact. The bi-modal unbounded distribution trained under the slippage objective produced the most efficient execution strategy by using its two modes to optimise impact and risk separately. The uni-modal distribution trained with the slippage function produces an execution strategy which can produce optimal VaR at a 95\% confidence level.

One area for future work is to incorporate a contextual bandit framework. This is where the RL agent learns a policy that maps context to an action to maximise a reward \cite{bouneffouf2020survey, cannelli2023hedging}. An agent trained in this manner could update its beliefs during execution, responding to the risk and impact of the execution in real time. Another area which may be interesting to explore is using the efficient frontier as the loss function. The distance to the frontier could be used as the loss function, which would align learning objectives more closely with economic theory \cite{hendricks2014reinforcement}.

Overall, this study positions RL not just as a viable tool, but as a compelling alternative to traditional algorithmic execution methods in financial markets. We expect extensions, set out above, to further justify the use of RL for optimising market execution, especially where considering correlated market movements and execution across different market regimes.

\section{Significance}
This work demonstrates the potential of RL agents in discovering efficient execution strategies under realistic market conditions. By leveraging a high-fidelity simulation environment, we show that RL agents can learn policies that balance market impact and risk more effectively than traditional strategies such as TWAP and VWAP.

Our approach highlights two key insights. First, RL agents can adaptively learn optimal trade timing by responding to the volume profile and intraday dynamics, resulting in lower transaction costs. Second, by modifying the reward objective to minimise slippage or impact, different execution profiles can be produced to suit varying risk preferences.

\bibliographystyle{ACM-Reference-Format}
\bibliography{bib} 


\end{document}